\documentclass[a4paper, amsfonts, amssymb, amsmath, reprint, showkeys,  twoside,twocolumn,superscript address]{revtex4-2}
\usepackage[english]{babel}
\usepackage[utf8]{inputenc}
\usepackage[colorinlistoftodos, color=green!40, prependcaption]{todonotes}
\usepackage{comment}
\usepackage{bbm}
\usepackage{amsthm}
\usepackage{mathtools}
\usepackage{physics}
\usepackage{xcolor}
\usepackage{graphicx}
\usepackage[normalem]{ulem}
\usepackage[left=23mm,right=13mm,top=35mm,columnsep=15pt]{geometry} 
\usepackage{adjustbox}
\usepackage{placeins}
\usepackage[T1]{fontenc}
\usepackage{lipsum}
\usepackage{csquotes}
\usepackage{xspace}
\newcommand{\vct}[1]{\mathbf{#1}}
\usepackage[pdftex, pdftitle={Article}, pdfauthor={Author}]{hyperref} 
\DeclareSymbolFont{bbgreek}{U}{bbold}{m}{n}
\DeclareMathSymbol{\bbmu}{\mathbb}{bbgreek}{'26}
\DeclareMathSymbol{\bbeps}{\mathbb}{bbgreek}{'17}

\begin{document}
\title{Equilibrium forces on non-reciprocal materials}

\author{David Gelbwaser-Klimovsky}
    \email[Correspondence email address: ]{dgelbi@technion.ac.il}
    \affiliation{Schulich Faculty of Chemistry and Helen Diller Quantum Center, Technion- Israel Institute of Technology, Haifa 3200003, Israel}
\author{Noah Graham}
\affiliation{Department of Physics, Middlebury College, Middlebury, VT 05753  USA}
\author{Mehran Kardar}
      \affiliation{Department of Physics, Massachusetts Institute of Technology, Cambridge, MA 02139, USA}
\author{Matthias Kr\"uger}
     \affiliation{
     Institute for Theoretical Physics, Georg-August-Universität, 37077 Göttingen, Germany
     }

\begin{abstract}
We discuss and analyze the properties of Casimir forces acting between nonreciprocal objects in thermal equilibrium. By starting from the fluctuation-dissipation theorem and splitting the force into those arising from individual sources, we show that if all temperatures are equal, the resulting force is reciprocal and is derivable as the gradient of a Casimir (free) energy. While the expression for the free energy is identical to the one for reciprocal objects, there are several distinct features: To leading order in reflections, the free energy can be decomposed as the sum of two terms, the first corresponding to two reciprocal objects, and the second corresponding to two anti-reciprocal objects. The  first term is negative and typically yields attraction, while the second can have either sign. For the case of two objects that are each other's mirror images, the second term is positive and yields repulsion. The sum of terms can lead to overall repulsive forces, in agreement with previous observations. Stable configurations, ruled out for reciprocal cases, appear possible for nonreciprocal objects. We show that for three objects, a three-body free energy exists, indicating that previously found persistent heat currents in situations of three objects cannot be used to produce  persistent torques.
\end{abstract}

\maketitle
Steady currents are commonly associated with the lack of thermal equilibrium.
The existence of steady currents indicates the absence of detailed balance \cite{zia2007probability}, which can have important consequences, such as violation of the fluctuation-dissipation theorem \cite{Agarwal72}, the presence of activity  \cite{platini2011measure,battle2016broken,gnesotto2018broken}, lasing without inversion, \cite{scully2010quantum} and the increase of power extraction in photocells \cite{scully2011quantum}. 

Recently it was shown that non-reciprocal materials can support steady heat currents in a system that is at thermal equilibrium \cite{zhu2016persistent,khandekar2019thermal} by breaking detailed balance. Examples of non-reciprocal materials include ferrimagnetic compounds, magnetized plasmas,  and space-time modulated media, among others \cite{caloz2018electromagnetic}.
Besides  heat currents,  fluctuation of charges and electric currents inside  non-reciprocal bodies produce Casimir forces that differ from those produced by reciprocal systems \cite{fuchs2017casimir}. 
Non-reciprocal systems have been shown to produce new features absent in their reciprocal counterparts, such as novel  lateral forces \cite{gelbwaser2021near,silveirinha2018fluctuation,giron2019lateral}, repulsive forces \cite{farias2020casimir,jiang2019chiral}, recoil forces \cite{PhysRevB.97.201108}, nontrivial
optical torques \cite{gangaraj2018optical,lindel2018inducing}, lateral thermal-fluctuations-induced forces \cite{khandekar2021nonequilibrium,tsurimaki2021casimir,khandekar2019thermal,nefedov2017lateral} and heat transfer with unique properties \cite{fan2020nonreciprocal,zhu2016persistent,ott2020anomalous,biehs2020near}. Furthermore, some theorems restricting the properties of forces are based on the assumption that the systems are reciprocal \cite{Rahi10,kenneth2006opposites}, indicating the possibility of finding other novel effects in non-reciprocal systems.

In this paper, we study the nature of  Casimir forces produced by systems that break detailed balance  at thermal equilibrium. We achieve this by analyzing the properties of forces produced by  non-reciprocal materials and test whether they are also restricted by standard force theorems. We show that these forces are conservative, forbidding the steady production of mechanical work at equilibrium and therefore preventing them from being used as a way to increase power extraction in photocells or other heat-engine like devices. Moreover, the conservative property of the equilibrium force puts into question the results presented in Ref.~\cite{nefedov2017lateral} on a lateral force in translationally invariant setups, as well as the experimental proposal for using forces to measure equilibrium persistent heat currents in translation invariant geometries proposed in Ref.~\cite{khandekar2019thermal}. As was recently shown, steady work extraction in non-reciprocal heat engines requires at least two different temperatures \cite{gelbwaser2021near,guo2021single}. Furthermore, we show that non-reciprocity may in principle allow for stable configurations, as the Laplacian of the free energy can be positive, in violation of Earnshaw's theorem, which reciprocal systems must fulfill \cite{Rahi10}. This comes together with the possibility of repulsion for non-reciprocal objects \cite{fuchs2017casimir}. Specifically, to leading order in reflections, the free energy is the sum of two terms, corresponding to two reciprocal and two anti-reciprocal objects, respectively. For two objects that are each other's mirror images, the latter can be shown to yield a positive free energy and a repulsive force. Repulsive Casimir forces are known to exist for bodies immersed in a dielectric \cite{Dzyaloshinskii61}, or for systems out of equilibrium  \cite{Henkel02,Cohen03,Antezza08,Bimonte11}.

\section{Casimir Free Energy}
\subsection{System and Setup}
We consider the situation depicted in Fig.~\ref{fig:bod}, which comprises a collection of bodies immersed in vacuum. The system is at thermal equilibrium, i.e., all bodies as well as the surroundings are at the same temperature $T$.

The objects are described by their classical scattering operators $\mathbb{T}=T_{ij}(\vct{r},\vct{r}')$ \cite{rahi2009scattering, kruger2012trace}, which are $3\times3$ matrices depending on two spatial arguments $\vct{r}$ and  $\vct{r}'$.  $\mathbb{T}_1$ denotes the operator for object 1, and $\mathbb{T}_{\bar 1}$ is the operator for the remaining objects. This notation allows us to treat the case of two objects as well as that of several objects.

For reciprocal bodies, micro-reversibility constrains $\mathbb{T}=\mathbb{T}^T$, where $[T_{ij}(\vct{r},\vct{r}')]^T=T_{ji}(\vct{r}',\vct{r})$ is the transpose, corresponding to switching spatial arguments and matrix indices. Here we consider objects that are allowed to be non-reciprocal \cite{asadchy2020tutorial,caloz2018electromagnetic}, so that in general, $\mathbb{T}\not=\mathbb{T}^T$.     
\subsection{Derivation of Casimir Free Energy from Fluctuation Dissipation Theorem}
\begin{figure}
    \centering
    \includegraphics[width=0.475\textwidth]{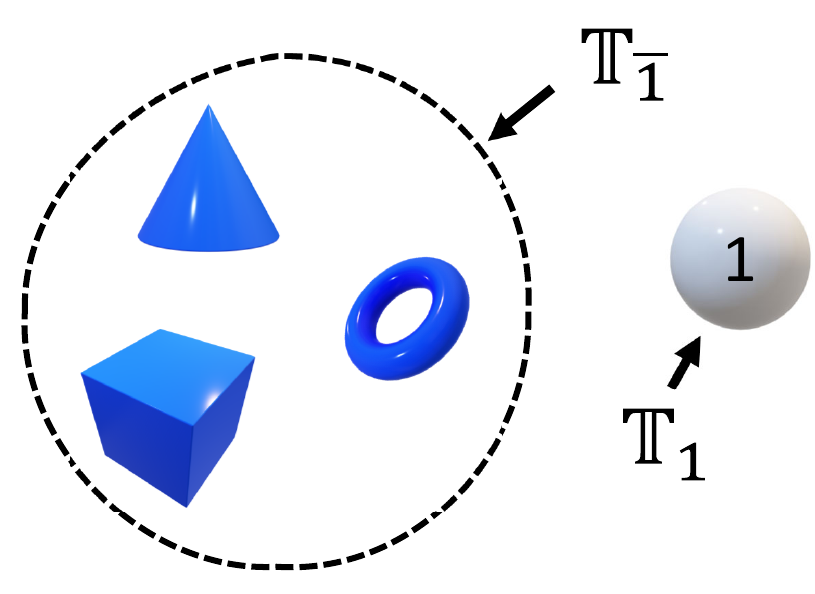}
    \caption{We consider multiple objects surrounded by vacuum, where all the objects and their surroundings are assumed to be at the same temperature T. We study the properties of the total force on object 1. }
    \label{fig:bod}
\end{figure}

As mentioned above, the systems depicted in Fig.~\ref{fig:bod}, may break time-reversal symmetry and also detailed balance \cite{zhu2016persistent}. It is thus not obvious how equilibrium Casimir forces can be computed \cite{fuchs2017casimir} and what  their properties are. We thus start by showing that Casimir forces among non-reciprocal or reciprocal bodies at thermal equilibrium are gradients of a potential energy, whose form is identical to the known case of reciprocal media \cite{rahi2009scattering}. This is in stark contrast with non-reciprocal forces \cite{ivlev2015statistical} and forces on active non-reciprocal mechanical systems \cite{scheibner2020odd} where the lack of thermal equilibrium allows the presence of non-conservative active forces. 

We therefore start from the corresponding expressions found for {\it non-equilibrium} forces \cite{kruger2012trace,gelbwaser2021near}, but then take all temperatures equal. This derivation is thus based on the assuming validity of the fluctuation dissipation theorem locally in each non-reciprocal system, which in turn is based on the existence of a Boltzmann distribution~\cite{Landauel}.

The force on $1$   is thus written as a sum of  three forces, corresponding to the sources in the different objects and the environment \cite{kruger2012trace}
\begin{align}
{\bf F}_{eq}^{(1)}={\bf F}_{\bar 1}^{(1)}+{\bf F}_{1}^{(1)}+{\bf F}_{env}^{(1)}.
\end{align}
${\bf F}_{\bar 1}^{(1)}$ is the force due to fluctuations sourced in objects  $\bar 1$, and reads
\begin{gather}
{\bf F}_{\bar 1}^{(1)}=\frac{\hbar}{\pi}\int_{0}^{\infty}d\omega\left[\frac{1}{e^{\frac{\hbar\omega}{k_{B}T}}-1}+\frac{1}{2}\right] \times \notag\\ \Re \Tr \left\{ \nabla(1+\mathbb{G}_{0}\mathbb{T}_{1})\mathbb{G}_{0}\mathbb{W}_{\bar 1,1}\mathbb{G}_{0}\times \phantom{\frac{1}{1}}\right.\notag\\
\left. \left[\frac{\mathbb{T}_{\bar 1}-\mathbb{T}_{\bar 1}^{\dagger}}{2i}-\mathbb{T}_{\bar 1}\Im[\mathbb{G}_{0}]\mathbb{T}_{\bar 1}^{\dagger}\right]\mathbb{G}_{0}^{*}\mathbb{W}_{\bar 1,1}^{\dagger}\mathbb{G}_{0}^{*}\mathbb{T}_{1}^{\dagger}\right\},\label{eq:F21}
\end{gather}

\noindent where $\mathbb{W}_{\bar 1,1}
=\mathbb{G}_0^{-1}\frac{1}{1-\mathbb{G}_0\mathbb{T}_{\bar 1}\mathbb{G}_0\mathbb{T}_1}$ is the multiple scattering operator and $\mathbb{G}_0$ is the free Green's function. Note that, in contrast to the formulas for nonequilibrium Casimir forces \cite{kruger2012trace}, we include here the zero-point term:
${\bf F}_{1}^{(1)}$   is the  ``self-force''  which is produced by the fluctuating charges in object $1$ itself,
\begin{gather}
{\bf F}_{1}^{(1)}=
\frac{\hbar}{\pi}\int_{0}^{\infty}d\omega\left[\frac{1}{e^{\frac{\hbar\omega}{k_{B}T}}-1}+\frac{1}{2}\right] \times \notag\\
\Re\Tr\left\{ \nabla (1+\mathbb{G}_{0}\mathbb{T}_{\bar{1}}) \mathbb{G}_{0}\mathbb{W}_{1,\bar 1}\mathbb{G}_{0}\times \phantom{\frac{1}{1}}
\right. \notag\\
\left.
\left[\frac{\mathbb{T}_{1}-\mathbb{T}_{1}^{\dagger}}{2i}-\mathbb{T}_{1}\Im[\mathbb{G}_{0}]\mathbb{T}_{1}^{\dagger}\right]\mathbb{G}_{0}^{*}\mathbb{W}_{1,\bar 1}^{\dagger}\right\}. \label{eq:F11}
\end{gather}
The self-force is zero for an isolated symmetric object, but potentially non-zero in the presence of other objects.
 The sources in the environment yield the force component ${\bf F}_{env}^{(1)}$, which is found to be
\begin{gather}
{\bf F}_{env}^{(1)}=\frac{\hbar}{\pi}\int_{0}^{\infty}d\omega\left[\frac{1}{e^{\frac{\hbar\omega}{k_{B}T}}-1}+\frac{1}{2}\right] \times \notag\\
 \Re\Tr\left\{ \nabla(1+\mathbb{G}_{0}\mathbb{T}_{1})\mathbb{G}_{0}W_{\bar 1,1}(1+\mathbb{G}_{0}\mathbb{T}_{\bar 1})\Im[\mathbb{G}_{0}] \times \right. \notag\\
\left.
(1+\mathbb{T}_{\bar 1}^{\dagger}\mathbb{G}_{0}^{*})\mathbb{W}_{\bar 1,1}^{\dagger}\mathbb{G}_{0}^{*}\mathbb{T}_{1}^{\dagger}\right\}.\label{eq:Fenv} \end{gather}
For reciprocal objects, the existence of a Casimir free energy for equilibrium systems is well established \cite{lifshitz1956zh,Neto08,Reid09, rahi2009scattering}. So, if $\mathbb{T}_i=\mathbb{T}_i^T$ for all $i$, Eqs.~\eqref{eq:F21}, \eqref{eq:F11} and \eqref{eq:Fenv} \emph{must} sum up to an expression that is the gradient of the known free energy \cite{rahi2009scattering}.  
Here we show this  for non-reciprocal objects. These pose additional challenges, because the operators to be traced do not enjoy the same symmetries as for reciprocal cases. For example, the term $\mathbb{T}_{1}\Im[\mathbb{G}_{0}]\mathbb{T}_{1}^{\dagger}$ is, for reciprocal cases, equal to the complex conjugate of $\mathbb{T}^\dagger_{1}\Im[\mathbb{G}_{0}]\mathbb{T}_{1}$. This property directly implies that heat transfer between reciprocal objects is symmetric \cite{kruger2012trace}. Because this property is absent for nonreciprocal objects, the corresponding  heat transfer is not necessarily symmetric \cite{herz2019green}. It is thus not obvious a priori  whether Eqs.~\eqref{eq:F21}, \eqref{eq:F11}, and \eqref{eq:Fenv} sum up to a gradient force for non-reciprocal cases.

After some algebraic manipulations detailed in SI, we find that the force is 
\footnote{We note a missing minus sign in Eq.\ (A6) of Ref.~\cite{kruger2012trace}}
\begin{gather}
{\bf F}_{eq}^{(1)}=-\nabla_{\mathcal{O}_{(1)}}\frac{\hbar}{\pi}\int_{0}^{\infty}d\omega\left[\frac{1}{e^{\frac{\hbar\omega}{k_{B}\mathbb{T}}}-1}+\frac{1}{2}\right] \times \notag\\
\Im\Tr\left[\log\left(1-\mathbb{G}_{0}\mathbb{T}_{1}\mathbb{G}_{0}\mathbb{T}_{\bar 1}\right)\right],\label{eq:fpot}
\end{gather}
 which is the derivative of a function with respect to the position of the object
$1$. This implies that the force is conservative. The function  inside the derivative is the free energy $\cal{F}$,
\begin{gather}
{\cal F}=\frac{\hbar}{\pi}\int_{0}^{\infty}d\omega\left[\frac{1}{e^{\frac{\hbar\omega}{k_{B}\mathbb{T}}}-1}+\frac{1}{2}\right] \times \notag\\
\Im\Tr\left[\log\left(1-\mathbb{G}_{0}\mathbb{T}_{1}\mathbb{G}_{0}\mathbb{T}_{\bar 1}\right)\right].\label{eq:fen}
\end{gather}

The conservative nature of the force rules out the possibility of building an engine at thermal equilibrium, as may be expected from thermodynamics: Extraction of mechanical work from a system at a single  temperature is prohibited. 
 In contrast, as shown in \cite{gelbwaser2021near}, for specific geometries non-reciprocity plays a key role for building a heat engine operating with bodies at different temperatures.
 
 In a translationally invariant setup, such as the non-reciprocal plate considered in \cite{khandekar2019thermal}, the free energy will be independent of the invariant coordinate. This implies that the force on objects in the vicinity of this plate is zero.
 
 Notably, the force in Eq.~\eqref{eq:fpot} and the free energy in Eq.~\eqref{eq:fen} are of identical form to the corresponding expressions for reciprocal objects \cite{rahi2009scattering, kruger2012trace}. Nevertheless, as we shall see, their properties can be quite different for the non-reciprocal case.
\subsection{Free energy as a Matsubara sum and large distance expansion}
To make contact with typical formulations of equilibrium Casimir forces, we rewrite Eq.~\eqref{eq:fen} in terms of a Matsubatra sum. We start by extending the integration range of Eq.~\eqref{eq:fen} to negative frequencies
\begin{align}
{\cal F}
&=\frac{\hbar}{4i\pi} \int_{-\infty}^\infty d\omega \coth\left[\frac{\hbar\omega\beta}{2}\right]\notag\\
&\mbox{Tr} \left\{ \log\left[1-\mathbb{G}_0\mathbb{T}_{1}\mathbb{G}_0\mathbb{T}_{\bar 1}\right]\right\}.
\end{align}
Because the free Green's function and the T operator are causal response functions, the trace is an analytic function in the upper complex plane \cite{Landauel} and the integral is performed via the residue theorem,
\begin{align}
{\cal F}&=k_BT\sum_{n=0}^\infty 
\mbox{Tr} \left\{ \log\left[1-\mathbb{G}_0\mathbb{T}_{ 1}\mathbb{G}_0\mathbb{T}_{\bar 1}\right]\right\},
\end{align}
where we sum over Matsubara frequencies $\omega\to i c \kappa_n=\frac{2\pi n i}{\hbar \beta}$. The term with $n=0$ is counted only half in the sum \cite{rahi2009scattering}, because the corresponding pole sits on the real frequency axis.

We will later investigate the so called one reflection approximation, where the log is expanded and the free energy is
\begin{align}
{\cal F}&\approx-k_BT\sum_{n=0}^\infty 
\mbox{Tr} \left\{ \mathbb{G}_0\mathbb{T}_{ 1}\mathbb{G}_0\mathbb{T}_{\bar 1}\right\}.\label{eq:pos}
\end{align}
This approximation becomes exact if the separation between objects is  large compared to their size, as studied below for two spherical particles. 
\section{properties for two objects}\label{sec:prop}
In this section,  we analyze the properties of the free energy and the  resulting forces for two objects. We use notation $\mathbb{T}_{\bar 1}=\mathbb{T}_{2}$. 
\subsection{Force Reciprocity}\label{sec:rec}
 Even though we consider non-reciprocal materials, the generated Casimir equilibrium forces are reciprocal in the sense that they obey Newton's third law. The scenario here is thus different from cases were non-reciprocal forces or interactions are found \cite{ivlev2015statistical}. This can be demonstrated by noticing that
\begin{gather}
 -\nabla_{\mathcal{O}_{1}}\Tr \left[\log\left(1-\mathbb{G}_{0}\mathbb{T}_{1}\mathbb{G}_{0}\mathbb{T}_{2}\right)\right]= \notag \\
 \Tr  \left[ \frac{1}{1-\mathbb{G}_{0}\mathbb{T}_{1}\mathbb{G}_{0}\mathbb{T}_{2}}\mathbb{G}_{0}
 \left(\nabla \mathbb{T}_1-\mathbb{T}_1\nabla \right) \mathbb{G}_{0}\mathbb{T}_{2}\right],\label{eq:recip}
\end{gather}
where we have used that $\nabla_{\mathcal{O}_1} \mathbb{T}_1=\nabla \mathbb{T}_1-\mathbb{T}_1\nabla $ and that we can pass the derivative through the free Green's function, as shown in Eqs.\ \eqref{eq:bef} and \eqref{eq:aft}.  Here $\nabla$ refers to the derivative acting to the right on the first argument of the subsequent expression.  Applying  $\nabla \mathbb{T}_2=\nabla_{\mathcal{O}_2} +\mathbb{T}_2\nabla$ to the last term and using the cyclic property of the trace we get that Eq.\ \eqref{eq:recip} is equal to
\begin{gather}
   - \Tr  \left[ \frac{1}{1-\mathbb{G}_{0}\mathbb{T}_{2}\mathbb{G}_{0}\mathbb{T}_{1}}\mathbb{G}_{0}
 \mathbb{T}_1\nabla_{\mathcal{O}_2}  \mathbb{G}_{0}\mathbb{T}_{2}\mathbb{G}_{0}\mathbb{T}_{1}\right] =\notag \\
 \nabla_{\mathcal{O}_2}  \Tr \left[\log\left(1-\mathbb{G}_{0}\mathbb{T}_{2}\mathbb{G}_{0}\mathbb{T}_{1}\right)\right].\label{eq:recip2}
\end{gather}
Combining \eqref{eq:recip} and \eqref{eq:recip2}, one obtains that the forces are reciprocal, that is  ${\bf F}^{(2)}_{eq}=-{\bf F}^{(1)}_{eq}$ independently of the bodies' reciprocity. This is in agreement with the expectation that nonreciprocal  forces require exchange of momentum with the environment, and can thus occur only in situations outside thermal equilibrium \cite{antezza2008casimir,Kruger11b}.

Generally, the statement of existence of a two-body potential already \emph{requires} force reciprocity: For two objects in free space, the free energy must be invariant under simultaneous translation of both objects. Hence it must be a function of $\boldsymbol{\mathcal{O}}_1-\boldsymbol{\mathcal{O}}_2$, and the forces derived from it must be reciprocal. We will below address the case of three bodies.     
\subsection{Only one  non-reciprocal object}

Assume that object 1 is non-reciprocal while  object 2 is reciprocal. In this case, using that $\Tr[\mathbb{O}]=\Tr[\mathbb{O}^T]$ for any operator $\mathbb{O}$, and using  the cyclic property of the trace, we obtain
\begin{align*}
    \mbox{Tr} \left\{ \mathbb{G}_0\mathbb{T}_1\mathbb{G}_0\mathbb{T}_{2}\right\}=\mbox{Tr} \left\{ \mathbb{G}_0\mathbb{T}^T_1\mathbb{G}_0\mathbb{T}_{2}\right\},
\end{align*}
or, rewriting,
\begin{align}
    \mbox{Tr} \left\{ \mathbb{G}_0(\mathbb{T}_1-\mathbb{T}^T_1)\mathbb{G}_0\mathbb{T}_{2}\right\}=0.\label{eq:tv}
\end{align}
We can thus write the contribution from the leading reflection, Eq.~\eqref{eq:pos}, as
\begin{align}
{\cal F}&=-\frac{1}{2}k_BT\sum_{n=0}^\infty 
\mbox{Tr} \left\{ \mathbb{G}_0\mathbb{T}_{2}\mathbb{G}_0(\mathbb{T}_1+\mathbb{T}^T_1)\right\}.\label{eq:pos4}
\end{align}
This means that the non-reciprocal part, represented by  $\mathbb{T}_1-\mathbb{T}^T_1$, does not contribute at one reflection to the free energy, if only a single object is non-reciprocal. For the limit of Eq.~\eqref{eq:pos}, the free energy will thus share all properties that are known for two reciprocal objects (see below). However, non-reciprocal properties enter at higher reflections, where the trace can be nonzero  because it may contain an even power of the asymmetric (non-reciprocal) operator $\mathbb{T}_1$ at higher orders. 
\subsection{Two distinct terms}
At one reflection, we can furthermore show that the force for two non-reciprocal objects is the sum of two distinct terms. We therefore write the scattering operators in terms of reciprocal and anti-reciprocal parts  
\begin{align*}
\mathbb{A}_i^{(\pm)}=\frac{1}{2}\left(\mathbb{T}_i\pm \mathbb{T}_i^T\right)=\frac{1}{2}\left(\mathbb{T}_i\pm \mathbb{T}_i^\dagger\right).    
\end{align*}
The last equality follows because $\mathbb{T}_i$ is real for imaginary frequency, since it is a real function of time $t$. For imaginary frequency, $\mathbb{A}_i^{(+)}$ is thus Hermitian, and $\mathbb{A}_i^{(-)}$ is anti-Hermitian. Noting from Eq.~\eqref{eq:tv} that there is no cross term between $\mathbb{A}_i^{(-)}$ and $\mathbb{A}_i^{(+)}$, the result for one reflection can be written
\begin{align}
   \notag{\cal F}&=-k_BT\sum_{n=0}^\infty \mbox{Tr} \left\{ \mathbb{G}_0\mathbb{T}_1\mathbb{G}_0\mathbb{T}_{2}\right\}=\\& -k_BT\sum_{n=0}^\infty  \biggl[\mbox{Tr} \notag\left\{\mathbb{G}_0\mathbb{A}_1^{(+)}\mathbb{G}_0\mathbb{A}_{2}^{(+)}\right\}\\&+ \mbox{Tr} \left\{\mathbb{G}_0\mathbb{A}_1^{(-)}\mathbb{G}_0\mathbb{A}_{2}^{(-)}\right\}\biggr].\label{eq:tterms}
   \end{align}
We see that, to leading order in the reflection expansion, the free energy is the sum of a free energy for two reciprocal objects and a free energy for two purely anti-reciprocal objects.   
We shall discuss their properties below.

\subsection{Energy sign and force direction} \label{sec:ensign}
The one-reflection approximation of Eq.~\eqref{eq:pos} allows the determination of the sign of $\cal F$.  
Starting with two reciprocal objects, it is useful to note, based on basic considerations of statistical physics, that $\mathbb{T}_{1}$ evaluated at imaginary frequencies is a non-negative symmetric operator  for non-magnetic objects in vacuum \cite{Landauel, kenneth2006opposites}. $\mathbb{G}_0\mathbb{T}_{2} \mathbb{G}_0$ is thus non-negative and symmetric as well (because $\mathbb{G}_0$ is symmetric and real). Because the product of two symmetric non-negative operators has a non-negative trace, we thus have 
\begin{align}
    \mbox{Tr} \left\{ \mathbb{G}_0\mathbb{T}_{2}\mathbb{G}_0\mathbb{T}_{1}\right\}\geq 0.\label{eq:nneg}
\end{align}
The free energy is thus a sum of non-positive terms, so that, for two reciprocal objects in leading order of scattering events,  
\begin{align}
{\cal F}\leq 0.
\end{align}
This implies that,  if the magnitude of ${\cal F}$ decreases with the objects separation, $d$, the force has to be attractive. 
This statement of attraction is in agreement with previous studies of forces between reciprocal bodies \cite{rahi2009scattering,kenneth2006opposites}. Below, we will also consider the case of two bodies which are each other's mirror images, in which case the force for reciprocal objects is attractive \cite{kenneth2006opposites}. 

For the non-reciprocal case, the $\mathbb{T}$ operators are not symmetric, and Eq.~\eqref{eq:nneg} does not hold. It is thus possible to obtain repulsive forces (even at small separations) with non-reciprocal media, as exemplified below and seen in \cite{fuchs2017casimir}. 

\subsection{Anti-reciprocal opposites repel}
This statement can be strengthened by considering two objects that are each other's mirror images. We follow \cite{kenneth2006opposites} to write $\mathbb{T}_{2}=\mathbb{J} \mathbb{T}_1  \mathbb{J}^\dagger$ with a unitary operator $\mathbb{J}$ that transforms between the original space and the mirror space. We also use the remarkable property \cite{kenneth2006opposites} that $\mathbb{G}_0\mathbb{J}$ is a positive operator. We thus write $\mathbb{G}_0\mathbb{J}=\mathbb{C}^\dagger\mathbb{C}=\mathbb{B}\mathbb{B}^\dagger$ without specifying $\mathbb{C}$ or $\mathbb{B}$. The two terms in \eqref{eq:tterms} are thus found to be 
\begin{align}
\notag\mbox{Tr} \left\{\mathbb{G}_0 \mathbb{A}_1^{(\pm)}\mathbb{G}_0\mathbb{A}_2^{(\pm)}\right\}&=\mbox{Tr} \left\{\mathbb{G}_0\mathbb{A}_1^{(\pm)}\mathbb{G}_0\mathbb{J}\mathbb{A}_1^{(\pm)}\mathbb{ J}^\dagger\right\}\\
\notag&=\mbox{Tr} \left\{\mathbb{C}\mathbb{A}_1^{(\pm)} \mathbb{B}\mathbb{B}^\dagger \mathbb{A}_1^{(\pm)}\mathbb{ C}^\dagger\right\}\\
&=\pm\mbox{Tr} \left\{(\mathbb{C}\mathbb{A}_1^{(\pm)} \mathbb{B})\left(\mathbb{C} \mathbb{A}_1^{(\pm)}\mathbb{ B}\right)^\dagger\right\}.\label{eq:rep}
\end{align}
Because
\begin{align}
\mbox{Tr} \left\{(\mathbb{C}\mathbb{A}_1^{(\pm)} \mathbb{B})\left(\mathbb{C} \mathbb{A}_1^{(\pm)}\mathbb{ B}\right)^\dagger\right\}\geq 0,
\end{align}
we have thus shown that to leading order in reflections, for two objects that are mirror images of each other the free energy is the sum of a negative term due to the reciprocal parts of the objects, and a positive term due to the anti-reciprocal part of the objects. 

Furthermore, we may also use that \cite{kenneth2006opposites} $\partial_a\mathbb{G}_0\mathbb{J}$ is a negative operator, i.e., $\partial_a\mathbb{G}_0\mathbb{J}=-\mathbb{D}^\dagger\mathbb{D}$, with another unspecified operator $\mathbb{D}$, and where $a$ is the distance between the mirror images \cite{kenneth2006opposites} along the mirror axis. Using this result in the same manner as above, we find that, for the term composed of two purely anti-reciprocal objects, 
\begin{equation}
\partial_a {\cal F}\leq 0.
\label{eq:repf}
\end{equation}
This means that the corresponding force is repulsive. We have thus shown that non-reciprocal contributions for mirror images lead to repulsive terms in the force.

Such repulsion and attraction between different types of objects is somewhat reminiscent of critical Casimir forces \cite{Hertlein08}.

\subsection{Stability}
Stability has been ruled out for equilibrium situations involving reciprocal objects \cite{Rahi10}.
 We repeat this calculation here, again resorting to the simpler case of the leading order in scattering events. We then investigate this question for non-reciprocal objects. We have for the Laplacian of the free energy   
\begin{align}
\nabla^2_{\mathcal{O}_1}{\cal F}(\mathcal{O}_1,\mathcal{O}_2)&=- k_BT\sum_{n=0}^\infty 
\mbox{Tr} \left\{ \mathbb{G}_0(\nabla^2_{\mathcal{O}_1}\mathbb{T}_1)\mathbb{G}_0\mathbb{T}_{2}\right\}.
\end{align}
We  use $\nabla_{\mathcal{O}_1} \mathbb{T}_1=\nabla \mathbb{T}_1-\mathbb{T}_1\nabla $, and obtain three terms,
\begin{gather}
\nabla^2_{\mathcal{O}_1}{\cal F}(\mathcal{O}_1,\mathcal{O}_2)=\notag\\- k_BT\sum_{n=0}^\infty \mbox{Tr} \biggl\{ \mathbb{G}_0[\nabla^2 \mathbb{T}_1-2\nabla\mathbb{T}_1\nabla+ \mathbb{T}_1\nabla^2]
\mathbb{G}_0\mathbb{T}_{2}\biggr\}.\label{eq:sign}
\end{gather}
We  use that, up to a $\delta$-function in space, $\nabla^2 \mathbb{G}_0=\frac{\kappa_n^2}{c^2}\mathbb{G}_0$ for imaginary frequency \cite{Rahi10}. This together with Eq.~\eqref{eq:nneg} shows that the last term is negative (including the overall minus sign). For the first term we use that the 
 operator $\nabla^2$ can be moved past the free Green's function, with the same conclusion.
The middle term appearing in Eq.~\eqref{eq:sign} is
\begin{align}
\mbox{Tr} \biggl\{ \mathbb{G}_0\nabla \mathbb{T}_1\nabla\mathbb{G}_0\mathbb{T}_{2}\biggr\}
=-\mbox{Tr} \biggl\{ \nabla \mathbb{T}_1\nabla^T\mathbb{G}_0\mathbb{T}_{2} \mathbb{G}_0\biggr\}\leq0,\label{eq:pos3}
\end{align}
where  $\nabla^T$ represents the derivative acting to the left on the second argument of the previous expression. By partial integration, we then have $\nabla=-\nabla^T$.
To establish the inequality, we have used that the trace of the product of the symmetric non-negative operators $\nabla \mathbb{T}_1\nabla^T$ and $\mathbb{G}_0\mathbb{T}_{2} \mathbb{G}_0$ must be non-negative. The Laplacian of the free energy with respect to the position of either of the particles is thus non-positive for reciprocal objects \cite{Rahi10}, and stability is ruled out.  

For non-reciprocal objects, we cannot rely on the above arguments, so the Laplacian cannot be proven to be non-positive. We will indeed provide a counterexample below, showing that the Laplacian can be positive and stable configurations with non-reciprocal systems appear possible in principle.

\section{Three bodies}
Eq.~\eqref{eq:fen} shows that the force acting on body 2 in the presence of several other objects is the gradient of a free energy. Together with the statements of Sec.~\ref{sec:rec}, the existence of a two-body potential for two objects is shown. However,  to our knowledge, this statement does  not imply that an $N$ body potential exists for $N>2$ objects. The case of three or more objects is especially interesting, since the persistent heat current discussed in Ref.~\cite{zhu2016persistent} requires three or more objects. It is thus worthwhile to test whether for three bodies, a three- body potential exists. We do this here to leading order in scattering reflections. We therefore introduce $\mathbb{T}_{23}$, the composite operator for objects $2$ and $3$. The force acting on object $1$ is then 
\begin{gather}
{\bf F}^{(1,eq)}=- \nabla_{\mathcal{O}_1}\frac{\hbar}{\pi} \int_0^\infty d\omega \left[\frac{1}{e^{\frac{\hbar\omega}{k_BT}}-1}+\frac{1}{2}\right] \times \notag \\
        \Im \mbox{Tr} \left\{\log\left[1-\mathbb{G}_0\mathbb{T}_1\mathbb{G}_0\mathbb{T}_{23}\right]\right\}.\label{eq:eqf2}
\end{gather}
We perform two series expansions. The first expands  the log
\begin{align}
\log\left[1-\mathbb{G}_0\mathbb{T}_1\mathbb{G}_0\mathbb{T}_{23}\right]= -\mathbb{G}_0\mathbb{T}_1\mathbb{G}_0\mathbb{T}_{23}+\dots.
\end{align}
The second expands the composite operator $\mathbb{T}_{23}$, 
\begin{align}
\mathbb{T}_{23}=\mathbb{T}_{2}+\mathbb{T}_{3}+\mathbb{T}_{2}\mathbb{G}_0\mathbb{T}_{3}+\mathbb{T}_{3}\mathbb{G}_0\mathbb{T}_{2}+\dots
\end{align}
With this expansion, we obtain the force acting on object 1, to leading order in scattering operators
\begin{gather}
        {\bf F}^{(1)}_{eq}={\bf F}_{2}^{(1,eq)}+{\bf F}_{3}^{(1,eq)}+ \notag \\ \nabla_{\mathcal{O}_1}\frac{\hbar}{\pi} \int_0^\infty d\omega \left[\frac{1}{e^{\frac{\hbar\omega}{k_BT}}-1}+\frac{1}{2}\right] \times \notag \\
        \Im \mbox{Tr} \left\{\mathbb{G}_0\mathbb{T}_1\mathbb{G}_0(\mathbb{T}_{2}\mathbb{G}_0 \mathbb{T}_{3}+\mathbb{T}_{3}\mathbb{G}_0 \mathbb{T}_{2})\right\},
\end{gather}
where ${\bf F}_{2(3)}^{(1,eq)}$ is the two-body force in the absence of object 3(2).
We thus can write
\begin{align}
        {\bf F}^{(1)}_{eq}=-\nabla_{\mathcal{O}_1}\mathcal{F}_1,
\end{align}
where
\begin{gather}
        \mathcal{F}_1= \mathcal{F}_{12} +  \mathcal{F}_{13}+\mathcal{F}_{23}- \notag \\
        \frac{\hbar}{\pi} \int_0^\infty d\omega \left[\frac{1}{e^{\frac{\hbar\omega}{k_BT}}-1}+\frac{1}{2}\right] \times \notag \\
        \Im \mbox{Tr} \left\{\mathbb{G}_0\mathbb{T}_1\mathbb{G}_0(\mathbb{T}_{2}\mathbb{G}_0 \mathbb{T}_{3}+\mathbb{T}_{3}\mathbb{G}_0 \mathbb{T}_{2})\right\}.
\end{gather}
\noindent Here $\mathcal{F}_{12}$ is the two-body free energy in absence of object 3. $\mathcal{F}_{23}$ can be added because it drops out when taking the derivative $\nabla_{\mathcal{O}_1}$.

The force on object 2 is found similarly,
\begin{align}
        {\bf F}^{(2)}_{eq}=-\nabla_{\mathcal{O}_2}\mathcal{F}_2,
\end{align}
with
\begin{gather}
        \mathcal{F}_2= \mathcal{F}_{12} +  \mathcal{F}_{13}+ \mathcal{F}_{23}-   \notag \\
        \frac{\hbar}{\pi} \int_0^\infty d\omega \left[\frac{1}{e^{\frac{\hbar\omega}{k_BT}}-1}+\frac{1}{2}\right]  \times \notag \\
        \Im \mbox{Tr} \left\{\mathbb{G}_0\mathbb{T}_2\mathbb{G}_0(\mathbb{T}_{1}\mathbb{G}_0 \mathbb{T}_{3}+\mathbb{T}_{3}\mathbb{G}_0 \mathbb{T}_{1})\right\}.
\end{gather}
The same can be done for object $3$.
Using the cyclic property of the trace, we can see that the forces are found as gradients of the same free energy, i.e.,
\begin{align}
        \mathcal{F}_3= \mathcal{F}_{2}=\mathcal{F}_{1}\equiv  \mathcal{F}(\mathcal{O}_1,\mathcal{O}_2,\mathcal{O}_3)
\end{align}
and the force on object $i$ is
\begin{align}
        {\bf F}^{(i)}_{eq}=-\nabla_{\mathcal{O}_i}\mathcal{F}(\mathcal{O}_1,\mathcal{O}_2,\mathcal{O}_3).
\end{align}
This demonstrates that, to leading order, a three-body potential $\mathcal{F}(\mathcal{O}_1,\mathcal{O}_2,\mathcal{O}_3)$ exists.

\section{Explicit examples}
In this section we discuss specific examples of the free energy for non-reciprocal objects.

 \begin{figure*}[htbp]
    \centering
    \includegraphics[width=\textwidth]{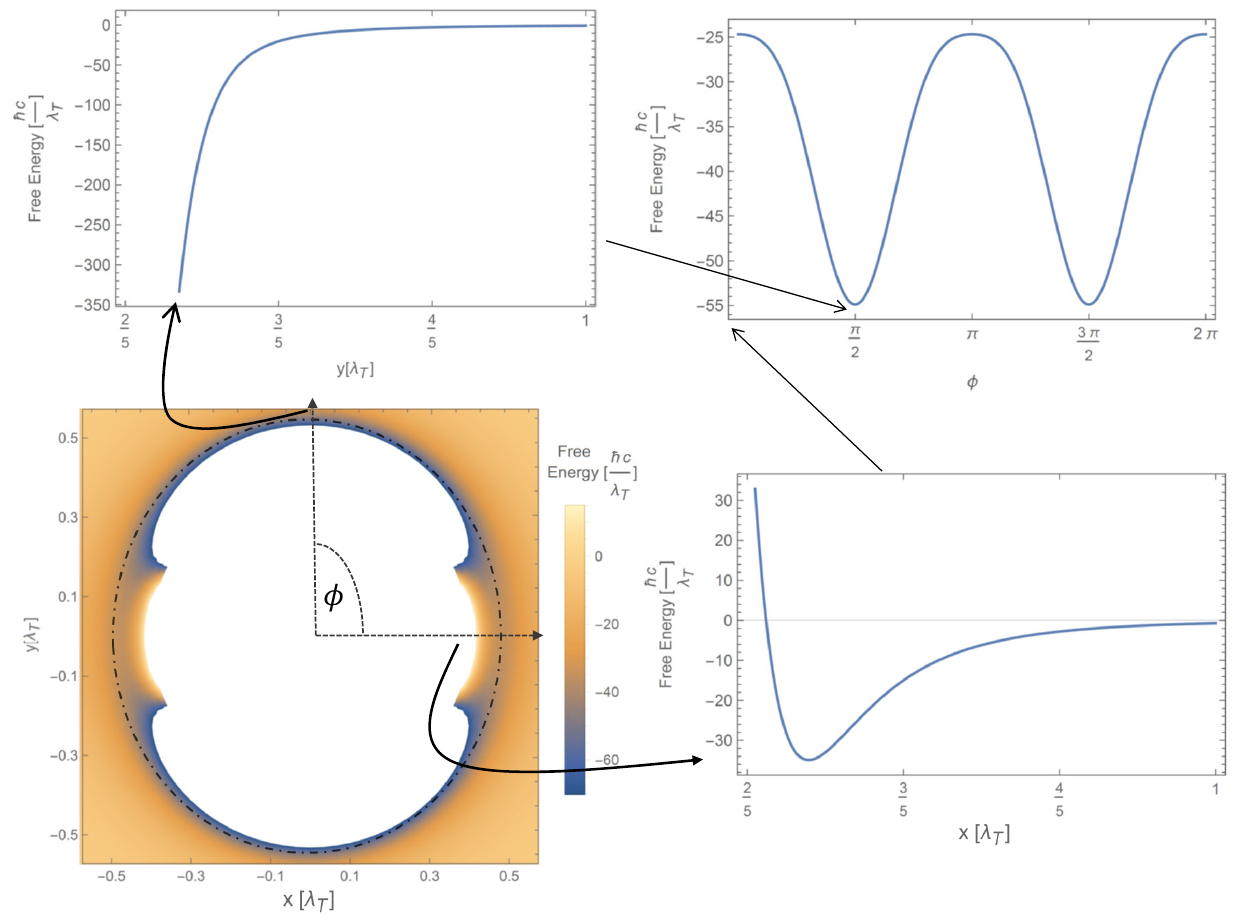}
    \caption{Free energy of particle 2 as function of its  position relative to particle 1 for a toy material. Both particles have same frequency-dependent polarizabilities (see Eq.\ \eqref{eq:permitf} and \cite{ott2020anomalous}), and the magnetic field points into the $x$ direction. Left-Bottom: Free energy in the $xy$ plane. Left-Top: Free energy as function of $y$ for $x=z=0$. Right-Bottom: Free energy as function of  $x$  for $y=z=0$ Right-Top: Free energy as function of the angle  $\phi$  for $z=0$ and radius $0.47\lambda_T$  (dashed circle in left-bottom figure). At this  separation, the free energy reaches a minimum along the $x-$axis, but it shows a maximum along $\phi$. Here $\omega_p=2\times10^8$ Hz, $\omega_{\tau}=2.46\times 10^{14}$ Hz, $\omega_B=1$ Hz and $T=300$ K.}
    \label{fig:real}
\end{figure*}

\begin{figure}
    \centering
    \includegraphics[width=0.5\textwidth]{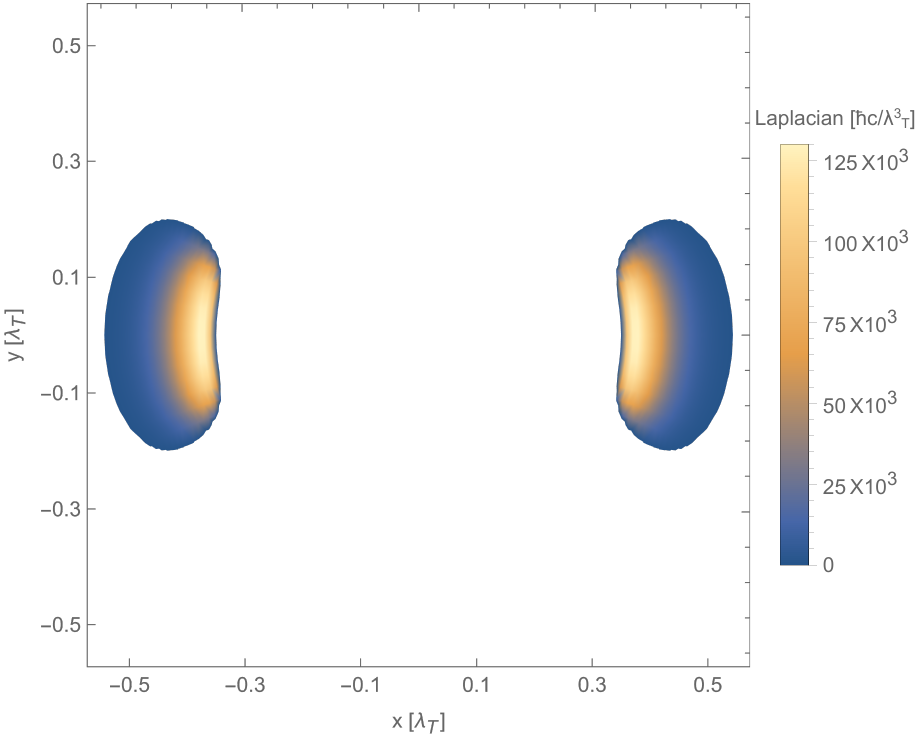}
    \caption{Laplacian of the free energy of particle 2 as function of its  position relative to particle 1. Only the positive regions are shown and they include the position of the minima in the radial direction ($z=0$, $x=0.47 \lambda_T$ and $y=0$), seen in Fig.~\ref{fig:real}. Same parameters as Fig.~\ref{fig:real}.}
    \label{fig:lap}
\end{figure}

\subsection{Toy model with frequency independent permittivity}
 A simple toy model allows to illustrate many of the above general findings, and to demonstrate the existence of positive Laplacian and repulsive forces.  

We start by considering two polarizable particles with separation $d$, which is large compared to the particles' sizes. To allow for analytic results, we assume that the particles' polarizability tensors are frequency-independent and have the following non-reciprocal form, with $\alpha_0$ (units of volume) and $b_i$ (dimensionless) real,
\begin{equation}
\alpha_i=\alpha_0 \left(\begin{array}{ccc}
1 & 0 & 0\\
0 & 1 & - b_i\\
0 & b_i &1
\end{array}\right).\label{eq:permit}
\end{equation}
This toy model, apart from frequency dependence, is of similar form to the response found for materials in the presence of an external magnetic field pointing in $x$-direction and with magnitude proportional to $b_i$. We will thus refer to $b$ also as a magnetic field in the following.

Taking one particle at the origin and the second at position $d(\cos\phi \sin \theta,\sin\phi \sin \theta,\cos \theta)^T$ we find for $d\gg\frac{\hbar c}{k_BT}$
\begin{gather}
 {\cal F} = \frac{k_B T\alpha_0^2}{32d^6\pi^2} \times \notag \\
\left[-12-5b_1b_2-3b_1b_2\left(\cos[2\theta]-2\cos[2\phi]\sin[\theta]^2\right) \right].\label{eq:long}
\end{gather}
For $b_i=0$ this agrees with literature \cite{rahi2009scattering} and the free energy is negative, so the force is attractive. Notably, there exists no term linear in $b_i$, in agreement  with Eq.~\eqref{eq:tterms}: A term linear in $b_i$ would result from a product of reciprocal and anti-reciprocal parts of $\alpha_i$, and is thus ruled out. 

The contribution from antireciprocal parts $\alpha_i-\alpha_i^T$ corresponds to terms quadratic in $b$. Indeed, these can be positive or negative (we will discuss the mirror symmetric case below). The free energy as a sum of both terms can indeed be positive, for example, for $\theta=\pi/2$, $\phi=0$ and $b_1b_2>3$. 

For short separation, $d\ll \frac{\hbar c}{k_BT}$, we find 
\begin{gather}
 {\cal F} =  \frac{\hbar c\alpha_0^2}{64d^7\pi^3} \times \notag \\
\left[-23-8b_1b_2-7b_1b_2\left(\cos[2\theta]-2\cos[2\phi]\sin[\theta]^2\right) \right].\label{eq:short}
\end{gather}
This limit shares the same general properties as the limit studied in Eq.~\eqref{eq:long}. Here,
for example, for $\theta=\pi/2$, $\phi=0$ and $b_1b_2>\frac{23}{13}$, the free energy is positive, and the force is repulsive. Recall that this is impossible for reciprocal particles. 

Continuing the example of $\theta=\pi/2$, $\phi=0$, the free energy is positive for short distance and negative at large distance for $\frac{23}{13}<b_1b_2<3$, which indicates the presence of a free energy minimum along the radial coordinate. Including displacements orthogonal to the radial direction, this point may be a minimum or a saddle point.  

Exploring this numerically, we find that the Laplacian at this point is positive, but, for the given choice of system, these are indeed saddle points. Below we discuss an  example of a frequency dependent permittivity.

Finally, we consider the case $\theta=0$ and $\phi=\frac{\pi}{2}$, i.e., where the two particles are displaced along the $z$ axis. They are then mirror images of each other, if we choose  $b\equiv b_2=-b_1$, meaning the magnetic field has to point in opposite directions. The free energy is then, for $d\gg\frac{\hbar c}{k_BT}$,
\begin{gather}
 {\cal F} = \frac{k_B T\alpha_0^2}{32d^6\pi^2} \left[-12+8b^2 \right],\label{eq:long2}
\end{gather}
obeying the properties found in Eq.~(\ref{eq:repf}). Similarly, for $d\ll \frac{\hbar c}{k_BT}$, we find
\begin{gather}
 {\cal F} =  \frac{\hbar c\alpha_0^2}{64d^7\pi^3} \left[-23+15b^2 \right].\label{eq:short2}
\end{gather}
Notably, if the particles are displaced along the $x$ axis ($\theta=\pi/2$ and $\phi=0$), mirror images are obtained without switching the direction of magnetic field, and the resulting anti-reciprocal part of free energy is positive in that case as well.

\subsection{Frequency-dependent permittivity} 

Turning to a more realistic case, we consider a frequency-dependent permittivity of the form of a standard  magneto-optical material. For a dc magnetic field pointing along the $x$ direction~\cite{ishimaru2017electromagnetic} it has the following form
\begin{equation}
\bbeps=\left(\begin{array}{ccc}
\epsilon_{p} & 0 & 0\\
0 & \epsilon_{d} & -i\epsilon_{f}\\
0 & i\epsilon_{f} & \epsilon_{d}
\end{array}\right),\label{eq:permitf}
\end{equation}
where  $\epsilon_d=1-\frac{\omega_p^2(1+\frac{i\omega_\tau}{\omega})}{(\omega+i\omega_\tau)^2-\omega_B^2}$, $\epsilon_p=1-\frac{\omega_p^2}{\omega(\omega+i\omega_\tau)}$ and $\epsilon_f=-\frac{\omega_B \omega_p^2}{\omega((\omega+i\omega_\tau)^2-\omega_B^2)}$ \cite{zhu2016persistent}.  Here, $\omega_p$ is the plasma frequency and $\omega_\tau$ describes relaxation effects; the non-reciprocity ($\epsilon_f\neq0$) due to the magnetic field is encoded via the cyclotron frequency $\omega_B$. We assume that both particles have the same permittivity, which we use to calculate the polarizability \cite{ott2020anomalous}. The parameters used are given in the caption of Fig.~\ref{fig:real}.

The numerical results are presented in Fig.~\ref{fig:real} and  Fig.~\ref{fig:lap} for a toy material. The parameters were chosen to make the effect easier to see in the plots. Further investigation is required to determine the presence of saddle points and the positivity of the Laplacian in realistic materials. Notably, the discussion of frequency independent cases is applicable to a large extent here as well. We see that repulsion is possible, and the free energy, as a function of particle distance can have a minimum as before. Also, here, these minima in the radial direction are {\it saddle points} in full 3D space. As a result while we showed in Sec.~\ref{sec:prop} that Earnshaw's theorem cannot be relied on for non-reciprocal media, this does not necessarily establish the existence of a minimum. Figure \ref{fig:lap} shows numerical results for the Laplacian of the free energy, displaying, as in the case of the frequency independent  polarizability, it is positive around the saddle point.

The quest for minima in this situation requires further investigation that could include the case of bodies with different permittivities or several bodies.

\section{summary}
Casimir forces acting between non-reciprocal materials in thermal equilibrium are conservative and reciprocal. The corresponding potential or free energy has a similar mathematical structure as the known one for reciprocal materials. Despite that, the properties of the free energy and the forces can be quite different between the reciprocal and non-reciprocal cases \cite{fuchs2017casimir,tsurimaki2021casimir}, which we have discussed in the so-called one-reflection approximation. In contrast to reciprocal cases, the free energy can be positive, and it can display minima. Furthermore, reciprocal and anti-reciprocal contributions yield two distinct terms in the free energy, which is thus a sum of the free energy between two reciprocal objects and the free energy between two anti-reciprocal objects. For two objects that are each other's mirror images, the second term is positive and the corresponding force is repulsive. These properties can be demonstrated in calculations for two non-reciprocal polarizable particles.

For three objects, we perform an expansion in reflections, finding that to leading order, a three-body potential exists. This case is of interest because persistent heat currents have been found before in situations involving three bodies. Our calculations -- to the given orders --  thus show that, despite existence of mentioned currents,  persistent torques are ruled out.

\section*{Acknowledgments}
D.\ G.-K.\ is supported by the Council for Higher Education Support Program for Hiring Outstanding Faculty Members in Quantum Science and Technology in Research Universities.
N.\ G.\ is supported in part by the National Science Foundation (NSF) by grant PHY-1820700.  M.\ K.\ is supported in part by NSF grant DMR-1708280.

\appendix
\section{Proof of Eq.\ \eqref{eq:fpot} }
 
 To obtain \eqref{eq:fpot} in the main text we use the following identities:
\begin{equation}
\mathbb{T}_{i}\mathbb{G}_{0}\mathbb{W}_{ji}=\mathbb{W}_{ij}\mathbb{G}_{0}\mathbb{T}_{i};	
\end{equation}

\begin{equation}
\mathbb{G}_{0}\mathbb{T}_{i}\mathbb{G}_{0}\mathbb{T}_{j}\mathbb{G}_{0}\mathbb{W}_{ij}=\mathbb{G}_{0}\mathbb{W}_{ij}\mathbb{G}_{0}\mathbb{T}_{i}\mathbb{G}_{0}\mathbb{T}_{j};
\end{equation}

\begin{equation}
\mathbb{G}_{0}\mathbb{T}_{i}\mathbb{G}_{0}\mathbb{T}_{j}\mathbb{G}_{0}\mathbb{W}_{ij}=\mathbb{G}_{0}\mathbb{W}_{ij}-1.\label{eq:id}.	
\end{equation}

 We get

\begin{gather}
{\bf F}_{eq}^{(1)}= \frac{\hbar}{\pi}\int_{0}^{\infty}d\omega\left[\frac{1}{e^{\frac{\hbar\omega}{k_{B}T}}-1}+\frac{1}{2}\right] \times \notag \\
\left[\frac{-1}{2i}\Re\Tr\left\{ \nabla \mathbb{G}_{0}^{\dagger}\mathbb{W}_{\bar 1 1}^{\dagger}\mathbb{G}_{0}^{*}\mathbb{T}_{1}^{\dagger}\right\} \right.   \notag\\
+\Re\Tr\left\{ \nabla(\mathbb{G}_{0}\mathbb{T}_{1})\mathbb{G}_{0}\mathbb{W}_{\bar 1 1}\frac{1}{2i}\right\} \notag\\ +\Re\Tr\left\{ \nabla \mathbb{G}_{0}\mathbb{W}_{\bar1 1}\frac{1}{2i}\right\} \notag\\
\left .-\frac{1}{2i}\Re\Tr\left\{ \nabla \mathbb{G}_{0}^{\dagger}\mathbb{W}_{1\bar1}^{\dagger}\right\}\right]. \label{eq:almost}
\end{gather}

 Taking the adjoint of the first term and using the cyclic properties
of the trace we get
\begin{equation}
-\frac{1}{2i}\Re\Tr\left\{\mathbb{G}_{0}\nabla \mathbb{T}_{1}\mathbb{G}_{0}\mathbb{W}_{\bar 1 1}\right\}.\label{eq:bef}
\end{equation}

We use that $\nabla_{r}\mathbb{G}_0(r-r')=-\nabla_{r'}\mathbb{G}(r-r')$. Furthermore, every operator product in the trace corresponds to the integral over a joint coordinate. We may thus use integration by parts (neglecting boundary terms), and  we get for the above term, 
\begin{equation}
-\frac{1}{2i}\Re\Tr\left\{\nabla \mathbb{G}_{0}\mathbb{T}_{1}\mathbb{G}_{0}\mathbb{W}_{\bar1 1}\right\}\label{eq:aft}
\end{equation}
and the first two terms of equation \eqref{eq:almost} cancel.

Next we calculate the adjoint of the last term of equation \eqref{eq:almost}
and using a similar procedure as to pass from Eq.\ \eqref{eq:bef} to
Eq \eqref{eq:aft} we get:

\begin{equation}
\frac{-1}{2i}\Re\Tr\left\{ \nabla \mathbb{G}_{0}^{\dagger}\mathbb{W}_{1\bar 1}^{\dagger}\right\} =\frac{-1}{2i}\Re\Tr\left\{ \nabla \mathbb{G}_{0}\mathbb{W}_{1\bar 1}\right\}. \label{eq:lastterm}
\end{equation}

Using identities \eqref{eq:id} we get

\begin{gather}
\frac{1}{2i}\Re\Tr\left\{ \nabla \mathbb{G}_{0}\mathbb{W}_{\bar 1 1}\right\} \notag \\ =\frac{1}{2i}\Re\Tr\{\nabla \mathbb{G}_{0}\mathbb{T}_{\bar 1}\mathbb{G}_{0}\mathbb{T}_{1}\mathbb{G}_{0}\mathbb{W}_{\bar1 1}\}= \notag \\
\frac{1}{2i}\Re\Tr\{\mathbb{T}_{1}\nabla \mathbb{G}_{0}\mathbb{T}_{\bar1}\mathbb{G}_{0}\mathbb{W}_{1 \bar 1}\mathbb{G}_{0}\}=\notag \\
\frac{1}{2i}\Re\Tr\{\nabla \mathbb{T}_{1} \mathbb{G}_{0}\mathbb{T}_{\bar 1}\mathbb{G}_{0}\mathbb{W}_{1 \bar 1}\mathbb{G}_{0}\} \notag \\
+\frac{1}{2i}\Re\Tr\{\nabla_{O_{1}}\mathbb{T}_{1} \mathbb{G}_{0}\mathbb{T}_{\bar 1}\mathbb{G}_{0}\mathbb{W}_{1 \bar 1}\mathbb{G}_{0}\},
\end{gather}
where in the last inequality we have used the identity 

\[
\mathbb{T}_{1}\nabla=\nabla \mathbb{T}_{1}+\nabla_{O_{1}}\mathbb{T}_{1}.
\]

We now apply identitites \eqref{eq:id} to \eqref{eq:lastterm} and
\eqref{eq:almost} 

\begin{gather}
-\Re \Tr \left\{ \nabla \mathbb{G}_{0}\mathbb{W}_{1 \bar 1}\frac{1}{2i}\right\} = \notag \\ 
-\Re\Tr\left\{ \nabla \mathbb{G}_{0}\mathbb{T}_{1}\mathbb{G}_{0}\mathbb{T}_{\bar 1}\mathbb{G}_{0}\mathbb{W}_{1 \bar 1}\frac{1}{2i}\right\}. 
\end{gather}

Putting everything together we get that

\begin{gather}
\frac{\hbar}{\pi}\int_{0}^{\infty}d\omega\left[\frac{1}{e^{\frac{\hbar\omega}{k_{B}T}}-1}+\frac{1}{2}\right] \times \notag\\ \frac{1}{2i}\Re\Tr\{\nabla_{O_{1}}\mathbb{G}_{0}\mathbb{T}_{1}\mathbb{G}_{0}\mathbb{T}_{ \bar 1}\mathbb{G}_{0}\mathbb{W}_{1 \bar 1}\}= \notag\\
\frac{\hbar}{\pi}\int_{0}^{\infty}d\omega\left[\frac{1}{e^{\frac{\hbar\omega}{k_{B}T}}-1}+\frac{1}{2}\right]  \times \notag\\ \frac{1}{2i}\Re\Tr\{\mathbb{G}_{0}\mathbb{W}_{1 \bar 1}\nabla_{O_{1}}\mathbb{G}_{0}\mathbb{T}_{1}\mathbb{G}_{0}\mathbb{T}_{ \bar 1}\}= \notag\\
-\frac{\hbar}{\pi}\int_{0}^{\infty}d\omega\left[\frac{1}{e^{\frac{\hbar\omega}{k_{B}T}}-1}+\frac{1}{2}\right]  \times \notag\\ \frac{1}{2i}\Re\Tr\nabla_{O_{1}}\log\left(1-\mathbb{G}_{0}\mathbb{T}_{1}\mathbb{G}_{0}\mathbb{T}_{ \bar 1}\right).
\end{gather}

\end{document}